\documentclass[aps,prl,reprint,superscriptaddress]{revtex4-2}

\usepackage{graphicx}
\usepackage{dcolumn}
\usepackage{bm}

\usepackage{xcolor}
\usepackage{ulem}
\usepackage{siunitx}
\usepackage[inkscapeformat=png]{svg}
\usepackage{soul}
\usepackage{amsmath} 

\begin{document}


\title{Membrane-mediated force transduction: Stick-slip motion of vesicles with fluid membranes}

\author{Paula Magrinya}
\affiliation{
Department of Theoretical Condensed Matter Physics, Condensed Matter Physics Center (IFIMAC) and Instituto Nicolás Cabrera, Universidad Autonoma de Madrid, 28049, Madrid, Spain
}
\author{Arin Escobar Ortiz}
\affiliation{
Department of Theoretical Condensed Matter Physics, Condensed Matter Physics Center (IFIMAC) and Instituto Nicolás Cabrera, Universidad Autonoma de Madrid, 28049, Madrid, Spain
}
\author{Juan L. Aragones}
\email{juan.aragones@uam.es}
\affiliation{
Department of Theoretical Condensed Matter Physics, Condensed Matter Physics Center (IFIMAC) and Instituto Nicolás Cabrera, Universidad Autonoma de Madrid, 28049, Madrid, Spain
}
\author{Laura R. Arriaga}
\email{laura.rodriguezarriga@uam.es}
\affiliation{
Department of Theoretical Condensed Matter Physics, Condensed Matter Physics Center (IFIMAC) and Instituto Nicolás Cabrera, Universidad Autonoma de Madrid, 28049, Madrid, Spain
}

\date{\today}

\begin{abstract}
How internal forces are transduced into motion through soft, fluid membranes remains a fundamental question in the study of active systems. To investigate this coupling, we develop a minimal system consisting of a single ferromagnetic particle encapsulated within a lipid vesicle with controlled membrane composition and phase behavior. An external rotating magnetic field actuates the particle, which rotates and translates along the inner membrane leaflet. This motion generates local slip in the membrane; near a substrate, the slip creates a shear gradient across the lubrication gap that propels the vesicle forward. Propulsion is intermittent and strongest when the particle moves near the vesicle bottom, where stress transmission is most effective. We find that the coupling between internal flows and vesicle motion is highly sensitive to membrane elasticity, excess area, and phase coexistence. Local membrane deformation and flow dissipate part of the stress, limiting the efficiency of force transduction. Additionally, membrane fluctuations and external boundaries reduce particle mobility, and in phase-separated membranes, line tension at domain boundaries deflects the particle and gradually reorients membrane structure. These results demonstrate that lipid membranes not only transmit internal stresses but also remodel themselves in response, actively shaping the dynamics of force transduction and motion in active systems.
\end{abstract}

\maketitle

The ability of cells to convert internal active forces into shape deformations and directed motion~\cite{janmey_dealing_2004} is central to diverse processes such as cell migration, endocytosis and mechanosensing~\cite{wang_transcriptional_2002,grodzinsky_cartilage_2000,bershadsky_crawling_2011,fu_mechanochemical_2023}. This force transduction occurs through the lipid membrane, a soft and fluid interface that plays an active mechanical role. Unlike solid boundaries, fluid membranes deform elastically in response to external or internal forces \cite{vutukuri_active_2020,takatori_active_2020,park_response_2022,le_nagard_encapsulated_2022,sciortino_active_2025}, or due to spontaneous thermal undulations---responses governed by membrane tension and bending rigidity~\cite{helfrich_steric_1978}. Phase coexistence further enriches their mechanics by introducing line tension at domain boundaries~\cite{esposito_flicker_2007}, while membrane viscosity allows flow~\cite{henle_hydrodynamics_2010,Oppenheimer_2021} and shear dissipation~\cite{amador_hydrodynamic_2021}. As a result, fluid membranes can redirect and transmit stresses tangentially despite lacking shear elasticity, making them active regulators of force transmission. As a minimal, controllable model system, giant unilamellar vesicles (GUVs) provide a platform to study membrane-mediated force transduction. When subjected to external shear or Poiseuille flows, the lipid membrane of GUVs exhibits characteristic deformations and tank-treading~\cite{vezy_adhesion_2007,abkarian_tank_2002,sturzenegger_membranes_2016,hamada_domain_2022,honerkamp-smith_membrane_2013,coupier_shape_2012, kraus_fluid_1996,gou_dynamics_2023}, reorganize membrane domains~\cite{sturzenegger_membranes_2016,hamada_domain_2022} and transduce external shear stresses into internal flows~\cite{honerkamp-smith_membrane_2013,amador_hydrodynamic_2021}. In contrast, how internally generated forces---such as cytoplasmic streaming driven by cortical motors in living cells~\cite{shelley_2024,dutta_self-organized_2024,kimura_endoplasmic-reticulum-mediated_2017,ganguly_cytoplasmic_2012,hueschen2024}---contribute to force transduction and motion is less understood. Though often associated solely with intracellular mixing~\cite{goldstein_physical_2015}, these internal flows likely play a more complex role in shaping membrane dynamics and motion. Addressing this open question requires synthetic systems with controllable internal activity. While theoretical models examine how internal rotating particles generate flows in confined geometries~\cite{aponte-rivera_simulation_2016,hoell_creeping_2019,kawakami_migration_2025}, and recent experiments with magnetic particles encapsulated in solid-like polymer vesicles demonstrate efficient force transmission through rigid membranes~\cite{magrinya_rolling_2024}, the coupling between internal activity and fluid lipid membranes remains largely unexplored~\cite{mateos-maroto_magnetic_2018}.

Here, to investigate how fluid membranes mediate the transduction of internal forces into whole-vesicle motion, we develop a minimal system consisting of a single ferromagnetic particle encapsulated within a lipid GUV produced by microfluidics. The particle is actuated by an external rotating magnetic field, generating internal fluid flows that interact with the deformable vesicle membrane. This setup allows us to examine not only how active stresses are transmitted to the membrane and drive motion, but also how the membrane mechanical properties---including fluidity, elasticity, excess area, and compositional heterogeneity---modulate, dissipate or redirect these stresses, revealing a dynamic interplay between internal activity and membrane response. 

\section*{Results and Discussion}

\subsection*{Rotating microparticles confined in vesicles}

To form vesicles with controlled size and membrane composition, each encapsulating a single ferromagnetic particle, we use water-in-oil-in-water (W/O/W) double emulsion drops with ultrathin oil shells as templates~\cite{arriaga_ultrathin_2014}. These templates are produced with a co-flow glass-capillary microfluidic device~\cite{kim_double-emulsion_2011}. The ultrathin thickness and chemical nature of the oil shell of the emulsions is critical for successful assembly of the lipid bilayer. Lipids are dissolved to a final concentration of 5 mg/mL in a mixture of 36 vol.\% chloroform and 64 vol.\% hexane. Evaporation of chloroform, which is more volatile and better solvent than hexane for lipids, favors their adsorption at the two O/W interfaces of the drop, while reducing the shell thickness. Excess lipids in the hexane-rich solvent mixture tend to aggregate inducing a depletion attraction between both monolayers and the ultimate assembly of the lipid membrane through solvent dewetting \cite{shum2011}. We study two different membrane compositions that produce fluid-like membranes, 100 mol\%  1,2-dioleoyl-sn-glycero-3-phosphocholine (DOPC) and a mixture containing 37.4 mol\% DOPC, 37.4 mol\% dipalmitoyl-sn-glycero-3-phosphocholine (DPPC), 25 mol\% cholesterol (Chol) and 0.2 mol\% rhodamine B 1,2-dihexadecanoyl-sn-glycero-3-phosphoethanolamine (DHPE-Rh), the later exhibiting liquid disordered (L$_{\mathrm{d}}$) - liquid ordered (L$_{\mathrm{o}}$) phase coexistence \cite{veatch_separation_2003}, with the L$_{\mathrm{d}}$ phase fluorescently labeled. The inner core of the vesicles results from encapsulation of an aqueous suspension of ferromagnetic particles (nominal radius, $R_p~=~$\SI{4}{\micro\meter}, carboxylic acid coating, CFM-80-5, Spherotech, Inc.) at a concentration of 1.5 mg/mL. This aqueous suspension additionally contains 2 wt.\% poly(vinyl alcohol) (PVA, 13000-23000 g/mol), which provides stability to the template until the bilayer is assembled. Additionally, it contains 8 wt.\% poly(ethylene glycol) (PEG, 6000 g/mol) or alternatively 200 mM of sucrose, as the latter enables the fabrication of smaller templates and thus smaller vesicles. The external medium containing the vesicles consists of an aqueous solution of either 200 mM sucrose or 220 mM glucose, depending on the internal vesicle composition, which matches the inner osmolarity of the vesicle to prevent mechanical stresses on the lipid membrane, simultaneously enabling vesicle sedimentation based on the density differences between the inner and external media. For observation, vesicles are transferred to a chamber made with two cover slips located approximately \SI{300}{\micro\meter} apart in a sandwich configuration. The bottom cover slip of the chamber is coated with bovine serum albumin (BSA) to prevent vesicle adhesion. 

\begin{figure}
\centering
\includegraphics[clip,scale=1,angle=0]{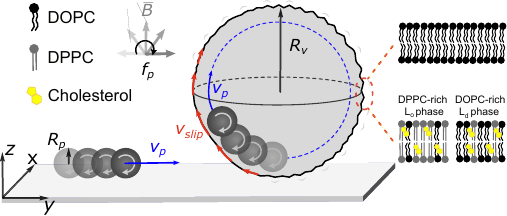}
\caption{Schematic representation of a particle rotating at a frequency $f_p$ around the $x$-axis. When free, the particle translates along the positive $y$-axis driven by the shear stress across the lubrication gap. However, when confined within a vesicle, it performs circular loops along the vesicle equator ($yz$-plane) in the direction opposite to the free particle due to an increased pressure force caused by membrane curvature, inducing a deformation and a slip velocity in the nearby vesicle membrane region, $v_{slip}$. The membrane can be composed either of pure DOPC or a ternary mixture of DOPC/DPPC/cholesterol, which phase-separates into microdomains with initial Janus morphology.}
\label{scheme}
\end{figure}

To drive particle rotation, we use an external rotating magnetic field of 10 mT with rotation axis parallel to the substrate ($x$-axis) and rotation frequency, $f_p$, varying from 1 to 10~Hz. Particle rotation creates a rotational flow in the fluid where it is suspended. In the lubrication limit, the presence of a limiting surface (the substrate) breaks the symmetry of the rotational flow, coupling the rotational and translational degrees of freedom of the particle. On the planar substrate, the no slip-boundary condition breaks the top-down symmetry of the flow field, creating a shear force that makes a clockwise rotating particle to translate along the positive $y$-axis or rolling direction \cite{goldmans_slow_1967,dou_programmable_2021,demirors_magnetic_2021,chamolly_irreversible_2020}, as illustrated schematically in the leftmost part of Fig.~\ref{scheme}. Importantly, this roto-translational coupling is directly influenced by the geometry of the limiting surface~\cite{gotze_relevance_2007,fang_magnetic_2020,bozuyuk_reduced_2022}. On a curved substrate, on the inner leaflet of the vesicle membrane, the particle follows circular loops along the vesicle equator ($zy$-plane) moving in the direction opposite to that of the free particle, as illustrated in Fig.~\ref{scheme} and Supplementary Movie (SM) 1. This sliding motion occurs due to the increased pressure force that results from breaking of the fore-aft symmetry of the rotational flow field on the curved limiting membrane \cite{magrinya_rolling_2024,mateos-maroto_magnetic_2018,gotze_flow_2010,caldag_numerical_2022}.

\subsection*{Flow transmission across fluid membranes}

\begin{figure}
\centering
\includegraphics[clip,scale=1,angle=0]{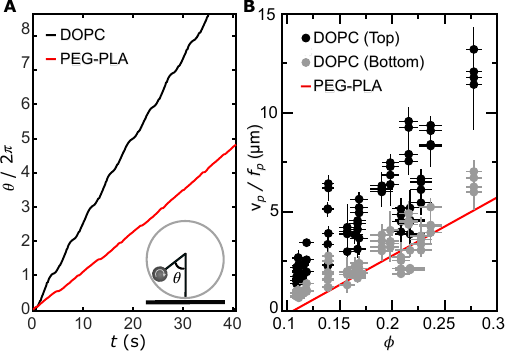}
\caption{A) Particle position within a fluid (DOPC) or a solid-like (PEG-PLA) vesicle as a function of time for a particle rotating at frequency $f_p=10$ Hz and degree of confinement $\phi=0.15$. B) Distance traveled by the particle per full rotation, $v_p/f_p$, along its circular loops as a function of the degree of confinement, $\phi$, when the particle is either at the top or bottom region of the DOPC vesicle. The red line is the case of the PEG-PLA vesicle~\cite{magrinya_rolling_2024}.}
\label{particle_vel}
\end{figure}

To determine the sliding velocity, $v_p$, of the particle on the inner leaflet of the lipid membrane, we track its position over time in bright-field microscope. Interestingly, in  stark contrast to polyethylene glycol-polylactic acid (PEG-PLA) polymer vesicles with solid-like membranes~\cite{magrinya_rolling_2024}, we observe that particle velocity varies along its circular trajectory, as exemplified in Fig.~\ref{particle_vel}A. In each loop along the circular trajectory, the particle slows down as it approaches the bottom of the vesicle, reaching its minimum velocity at this position. It then increases its velocity, reaching the maximum value at the top of the vesicle, repeating this cycle in successive loops.

To characterize the effect of confinement, defined as $\phi=R_p/R_v$, on the dynamics of the particle, we measure the sliding velocity $v_p$ at the top and bottom of the vesicle and normalize it by the particle rotational frequency. This normalization yields the distance traveled by the particle along the membrane per full rotation. Remarkably, at the bottom of the vesicle, this distance increases linearly with $\phi$ and is slightly larger than the values reported for solid-like vesicles~\cite{magrinya_rolling_2024}, as shown by the gray symbols lying just above the red line in Fig.~\ref{particle_vel}B. Moreover, at the top of the vesicle, the distance traveled per rotation is consistently larger than both at the bottom and in the solid case, with a noticeable steeper linear increase with $\phi$, as shown by the black symbols in Fig.~\ref{particle_vel}B. This distinct particle dynamics at the top and bottom of the vesicle suggests that the interplay between shear and pressure varies along the particle trajectory. In contrast to a solid boundary, a fluid lipid bilayer allows for flow transmission across the bilayer \cite{honerkamp-smith_membrane_2013,amador_hydrodynamic_2021}. At the top of the vesicle, where the outer fluid is unbounded, the flow generated by particle rotation dissipates, resulting in minimal shear stress and a higher sliding velocity. However, as the particle approaches the vesicle bottom, the flow transmitted across the membrane encounters the solid substrate beneath the vesicle. This additional hydrodynamic boundary enhances the local shear stress acting on the particle, thereby reducing its sliding velocity \cite{aponte-rivera_simulation_2016,hoell_creeping_2019}. However, the local shear stress on the particle at the bottom of fluid-like vesicles remains smaller than in solid-like vesicles~\cite{magrinya_rolling_2024}, because in the fluid case, the substrate lies beneath the membrane, whereas in the solid-like case, the rigid membrane itself act as the substrate. This additional separation in the fluid vesicle reduces the shear transmission to the particle, allowing pressure forces to play a more predominant role in the dynamics of the rotating particle.

\subsection*{Interplay between membrane shape fluctuations and particle dynamics}

\begin{figure}
\centering
\includegraphics[clip,scale=1,angle=0]{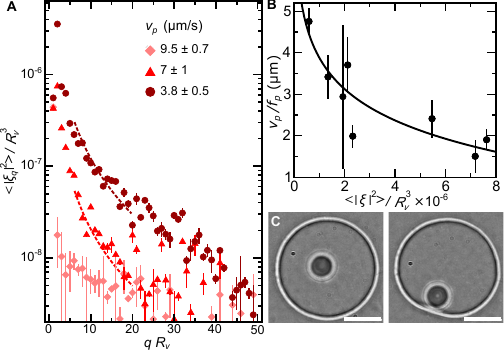}
\caption{A) Variation of the fluctuation amplitude, $\langle|\xi_q|^2\rangle$, as a function of the wavevector $q$ for a vesicle of radius, $R_v$, that undergoes osmotic deflation. The different colors correspond to different particle sliding velocities. Dashed lines correspond to fits using the Helfrich spectrum (See Methods and equation \ref{fit}) for membrane undulations \cite{helfrich_steric_1978,pecreaux_refined_2004}, yielding a bending modulus of $\kappa = 7.6\,k_B T$, and membrane tensions of $\sigma = 2.57\cdot10^{-6}$ N$\cdot$m$^{-1}$(triangles) and $\sigma = 1.14\cdot10^{-6}$ N$\cdot$m$^{-1}$ (circles). B) Variation of the distance traveled by the confined particle per full rotation as a function of the fluctuation amplitude. The black line is a fit to a logarithmic dependence of $v_p/f_p\propto\log{(1/\langle|\xi|^2\rangle)}$. C) Bright field microscope images showing visible shape deformations in a deflated vesicle as the particle approaches the vesicle membrane. Scale bar \SI{25}{\micro\meter}.}
\label{flickering}
\end{figure}

To understand how membrane mechanics influence particle dynamics, we examine the interaction between the encapsulated particle---sliding along the inner membrane leaflet--- and the vesicle membrane under varying tension conditions. Specifically, we induce excess membrane area in the vesicles by allowing evaporation of the outer medium, which concentrates the external solution and causes the vesicle to deflate. This deflation leads to a reduction in membrane tension and the emergence of shape fluctuations, providing a platform to probe their impact on particle sliding. To characterize membrane shape fluctuation during the deflation process, we perform flickering spectroscopy prior to driving particle motion. Once the fluctuation spectra for each tension state is recorded, we apply the magnetic field to drive particle rotation and measure its sliding velocity. The particle sliding velocity decreases by up to 40\% with increasing fluctuation amplitude, as shown in Fig.~\ref{flickering}A, where each curve is color-coded according to its associated particle velocity to provide a direct correlation between fluctuation amplitude and sliding behavior.

To further explore the link between membrane fluctuations and sliding dynamics, we examine how the fluctuation amplitude $\langle|\xi|^2\rangle=\sum_q{\langle|\xi_q|^2\rangle}$ correlates with the distance traveled by the particle in each full rotation, $v_p/f_p$. By analyzing a range of vesicles with varying tension states, we find that this distance systematically decreases as fluctuation amplitudes increase, as shown in Fig.~\ref{flickering}B. This trend is consistent with the particle moving closer to the inner membrane leaflet for smaller fluctuation amplitudes, where viscous coupling between particle rotation and translation become more effective. In contrast, large membrane fluctuations generate a repulsive pressure that increases the separation distance between the particle and the membrane, thereby weakening the hydrodynamic coupling \cite{aponte-rivera_simulation_2016}. This repulsion, known as the undulation pressure, arises from the entropic confinement of the shape fluctuations of a membrane as it approaches another surface and follows a power-law dependence with the distance to the confining substrate $P(d)\approx (k_BT)^2/\kappa_b d^3$ \cite{helfrich_steric_1978}, where $k_B$ is the Boltzmann constant, $T$ is the temperature, $\kappa_b$ is the membrane bending modulus and $d$ is the separation distance between the inner membrane leaflet and the surface of the particle in our particular case. Considering that this distance is set by the fluctuation amplitude, $d\propto \langle|\xi|^2\rangle^{1/3}$, we find the expected logarithmic dependence between particle velocity and particle-membrane separation distance \cite{goldmans_slow_1967,caldag_numerical_2022}, as shown by the black line in Fig.~\ref{flickering}B. 

To explore how the motion of the particle within the vesicles influences membrane shape, we analyze membrane deformations during particle sliding and find that the membrane becomes visibly disturbed, specially when the excess area is sufficiently large, as shown in Fig.~\ref{flickering}C. Notably, the fluctuation spectrum is only perturbed at long wavelengths, $qR_v<4$ (Fig. S1), suggesting that particle dynamics modulates the apparent membrane tension, without affecting membrane bending rigidity. In this tension-dominated regime, the membrane responds dynamically to the active hydrodynamic stresses imposed by the particle, redistributing its excess area through local shape deformations.

\subsection*{Interplay between membrane phase organization and particle motion}

\begin{figure*}
\centering
\includegraphics[clip,scale=1,angle=0]{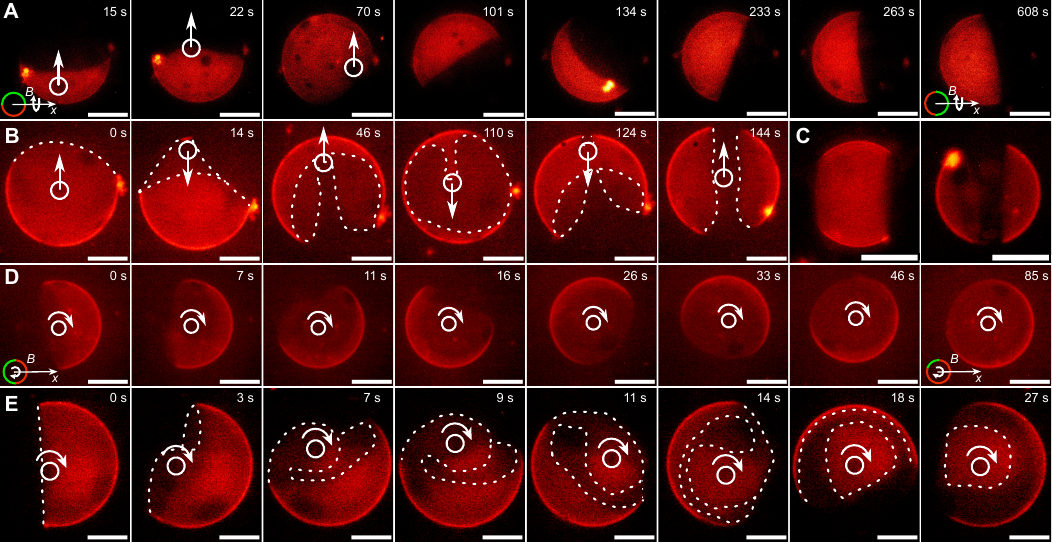}
\caption{Epi-fluorescence microscope images of a phase-separated vesicle, with the L$_{\mathrm{d}}$ phase labeled in red and the encapsulated particle, slightly visible as a dark spot highlighted as a white circle for visualization purposes. Particle rotating around the $x$-axis induces: A) domain alignment perpendicular to the $x$-axis when $f_p = 2$ Hz or B) a one-phase belt at the equator when $f_p = 5$ Hz. C) The steady-state configuration of B) may have the equator composed of either L$_{\mathrm{d}}$ (right) or L$_{\mathrm{o}}$ (left) phase. When the encapsulated particle spins around the $z$-axis induces: D) domain alignment perpendicular to the $z$-axis when $f_p = 5$ Hz or B) a one-phase belt at the equator when $f_p = 10$ Hz. Scale bar \SI{25}{\micro\meter}.}
\label{domains}
\end{figure*}

To further investigate the reciprocal coupling between membrane mechanics and particle dynamics, we study vesicle with phase coexistence, initially exhibiting a Janus configuration with two distinct membrane domains that differ in their mechanical properties. In contrast to single-phase membranes, the presence of coexisting domains introduces spatial variations in mechanical properties that may further modulate the local dynamics of the encapsulated particle. Interestingly, however, we find the same dependence of $v_p/f_p$ on $\phi$ as in single fluid membranes (Fig. S2), suggesting that differences in membrane mechanics between L$_{\mathrm{d}}$ and L$_{\mathrm{o}}$ domains play a minor role. Importantly, we observe off-trajectory displacements as particles encounter the domain boundary (SM2), which points to line tension as a key factor influencing particle dynamics in phase-separated membranes. This effect is most pronounced at low rotational frequencies ($f_p\leq2$~Hz), where the particle remains confined to a single domain and its motion is visibly constrained by membrane phase organization. Line tension arises from a local mismatch in membrane packing and generates asymmetric shear stresses that deflect the particle tangentially along the domain boundary. As a reactive response, in many cases, the initially tilted boundary gradually aligns with the vesicle equator, as shown in Fig.~\ref{domains}A and SM3, which allows the particle to recover circular looping trajectories at the equator and avoid diagonal drifts. This reorientation indicates that while line tension affects particle trajectory at low frequencies, the coupling is already bidirectional, with persistent sliding reorienting the domains as a result of the stress exerted by the particle. 

At higher rotational frequencies ($f_p>2$~Hz), the particle begins to actively remodel the membrane phase structure. The initial Janus configuration is progressively transformed into a three-domain structure, stabilized by the continuous sliding of the particle along the vesicle equatorial plane, as shown in Fig.~\ref{domains}B and SM4. The hydrodynamic force exerted by a particle of $R_p=$~\SI{4}{\micro\meter} rotating at $f_p=10$~Hz on a membrane with viscosity $\eta_m=10^{-9}$ Pa$\cdot$s$\cdot$m~\cite{Faizi2022} is $F\approx\eta_m v_{slip}\approx \eta_m 2\pi f_p R_p = 0.25$ pN, where $v_{slip}$ is the tangential slip velocity of the membrane region adjacent to the rotating particle, reflecting the local flow transmitted to the membrane by the particle. This force is larger than the line tension measured using domain boundary flicker spectroscopy, $\lambda\approx0.1$ pN~\cite{esposito_flicker_2007}. Therefore, in these conditions, the particle gains enough momentum to cross the boundary and deform it, in contrast to the lower frequency case, where $F<\lambda$ (Fig.~\ref{domains}A). Interestingly, similar reorganized structures are observed regardless of wether the L$_{\mathrm{o}} $ or L$_{\mathrm{d}}$ domains occupy the equatorial region, as shown Fig.~\ref{domains}C. This further confirms that the key factor governing membrane phase reorganization during particle sliding is the presence of line tension rather than the specific mechanical properties of each fluid domain. Phase organization diagrams showing the behavior of the domains---either aligning or forming a one-phase equatorial road--- as a function of $f_p$ and $\phi$, can be found in SI Appendix, Fig. S3.

To investigate the generality of the particle induced domain reorganization, we perform experiments using an alternative actuation mode where we apply the magnetic field around the $z$-axis, perpendicular to the substrate, making the particle spin in place at the vesicle bottom. In this configuration, for low frequencies ($f_p\leq2$~Hz), the domain boundary aligns perpendicular to the axis of rotation as in the sliding case, as shown in Fig.~\ref{domains}D. In contrast, when increasing the particle spinning frequency, the particle initially rotates directly at the domain boundary, gradually deforming it until the interfaces closes in on itself, creating the same type of bell observed in the sliding case. However, unlike in the sliding mode, the particle remains spinning outside the bell, confined within the small domain created, as shown in Fig.~\ref{domains}E. These complementary observations reinforce that membrane-phase reorganization can be robustly driven by hydrodynamic activity alone, independent of the specific mode of particle actuation.

The emergence of a three domain configuration provides two-domain boundaries that effectively stabilize the circular trajectory of the particle along the vesicle equator. In single-phase vesicles, wether fluid or solid-like, the particle typically deviates from the equatorial plane and spirals towards one of the vesicle poles (See SM1). In solid-like membranes, this behavior is attributed to the formation of an asymmetric Hill vortex that exerts a tangential force on the particle until the vortex dissipates \cite{magrinya_rolling_2024}. While the drift velocity toward the pole, $v_{pole}$, remains comparable between fluid and solid-like membranes (See SI Appendix, Fig. S4), the likelihood of reaching the pole is significantly reduced in fluid-like membranes: only 35\% in DOPC vesicles compared to 70\% in PEG-PLA vesicles. In most cases, the particles confined in fluid-like membranes follow spiraling trajectories that settle into steady-state orbits at intermediate latitudes (SM5). This reduced poleward displacement likely reflects a different internal flow structure within fluid vesicles. In contrast, the introduction of two domain boundaries in membranes with fluid coexistence stabilize the equatorial trajectory due to the hydrodynamic coupling with line tension, counteracting the drift observed in single-phase vesicles.

\subsection*{Asymmetric membrane slip and stick-slip vesicle translation}

\begin{figure}[t]
\centering
\includegraphics[clip,scale=1,angle=0]{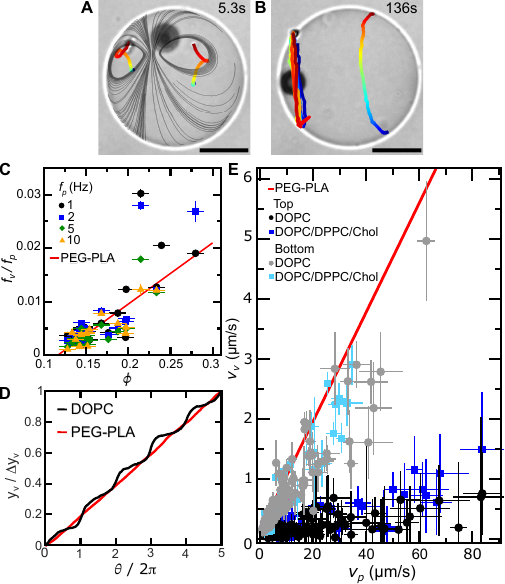}
\caption{ A-B) Trajectories of two membrane defects when the encapsulated particles is A) at the equator or B) at the pole. Gray lines in A) correspond to the theoretical streamlines expected for a point force on the membrane~\cite{Oppenheimer_2021}. Scale bar \SI{25}{\micro\meter}. C) Rotation frequency of the membrane at the equatorial region as function of degree of confinement for phase-separated vesicles. D) Evolution of the vesicle center of mass as a function of the angular position of the particle within the vesicle for both PEG-PLA and DOPC membranes. E) Translational velocity of vesicles as a function of particle velocity and position within the vesicle (top, bottom) for DOPC, and DOPC/DPPC/Chol membranes. The red line corresponds to solid-like PEG-PLA vesicles~\cite{magrinya_rolling_2024}.}
\label{vel_ves}
\end{figure}

To explore how internal flow couples to membrane motion, we track membrane defects during particle sliding. When the particle slides along the vesicle equator, membrane defects located symmetrically with respect to the equatorial plane trace vortex-like trajectories, as shown in Fig.~\ref{vel_ves}A and SM6. This is consistent with the dipolar flow field expected from a tangential point force applied to the membrane (see Appendix, Fig S5)~\cite{henle_hydrodynamics_2010,Oppenheimer_2021}. By contrast, when the particle rotates at the vesicle pole, membrane defects trace circular trajectories parallel to the equator, as shown in Fig.~\ref{vel_ves}B and SM6, with a rotation frequency that decreases with increasing defect-particle distance. This behavior is consistent with the streamline patterns generated by a point torque applied to the membrane, which produces concentric circular flows around the axis of rotation, with angular velocity decreasing with radial distance~\cite{henle_hydrodynamics_2010,Oppenheimer_2021}. These observations confirm that membrane motion in vesicles with fluid membranes arises from local membrane flow rather than from rigid-body rotation. In fact, rigid-body rotation is only expected when the vesicle radius is smaller than the Saffman length, $L_s=\eta_m/2\eta$, which defines the length scale beyond which momentum dissipation into the surrounding fluid dominates over viscous shear within the membrane~\cite{henle_hydrodynamics_2010,Oppenheimer_2021}.

To characterize the effect of confinement in the coupling between internal and membrane flows, we measure the rotational frequency of the vesicle, $f_v$, at the equatorial plane, by tracking the domain boundary in Janus vesicles (Fig.~\ref{domains}D and E), while driving particle spinning around the the $z$-axis at the vesicle bottom. Unlike tracking of membrane defects during particle sliding, this method allows us to probe the membrane flow field at a controlled radial distance from the point of torque application. The rotation frequency obtained by this method, and its dependence on $\phi$, closely resembles that of vesicles with solid-like membranes \cite{magrinya_rolling_2024}, as shown in  Fig.~\ref{vel_ves}C. This similarity reflects that the net torque is comparable in both cases.

To investigate how fluid membranes mediate force transduction during vesicle translation along the substrate, we track the center of mass of the vesicle along the positive $y$-axis. We observe that vesicle displacements are more pronounced when the sliding particle occupies the bottom region of the vesicle compared to when it occupies the top region, as shown by the black line in Fig. \ref{vel_ves}D. The origin of this stick-slip motion lies in the inner flow structure and viscous dissipation within a fluid membrane. Because the flow generated by particle rotation applies a point force on the closest membrane region that decays as $1/r$ (see Appendix, Fig S5), the local membrane velocity $v_{slip}^{bottom}$ when the particle is at the top must be smaller than that when the particle is at the bottom. In addition, since there is not rigid-body membrane rotation, the circulation of the fluid velocity field along any inner closed loop over the vesicle surface must be zero and thus, the net vorticity of the inner flows should vanish. This constraint results in the formation of an internal counterrotating vortex at the vesicle bottom when the particle is at the top, which opposes the field generated by the point force applied by the rotating particle, resulting in a nearly zero slip velocity in the membrane region near the substrate ($v_{\text{slip}}^{\text{bottom}} \approx 0$). Consequently, when the particle rotates at the top of the vesicle, there is negligible shear stress across the lubrication gap between the vesicle and the substrate, thereby producing no net vesicle translation (stick). In good agreement with this scenario, we measure a very small $v_v$ in single-phase and two-phase fluid vesicles independently on particle sliding velocity, as shown by the black circles and blue squares in Fig.~\ref{vel_ves}E. By contrast, when the particle is at the vesicle bottom, the local membrane slip velocity increases up to $v_{slip}^{bottom}\approx\eta_m2\pi f_p R_p$, generating a shear stress across the gap, that enables vesicle translation (slip). In good agreement with this scenario, we observe that $v_v$ increases linearly with $v_p$ ($v_p \propto f_p$), as shown by the gray circles and cyan squares in Fig.~\ref{vel_ves}E.

This behavior contrasts sharply with that of vesicles with solid-like membranes, which translate at constant velocity regardless of particle position due to the rigid-body rotation of the vesicle membrane (red line, Fig.~\ref{vel_ves}D)~\cite{magrinya_rolling_2024}. Moreover, the translational velocity of vesicles with fluid-like membranes when the particle is at the bottom is slightly lower than that of vesicles with solid-like membranes, as shown by the symbols lying just below the red line in Fig.~\ref{vel_ves}E. This reduced velocity reflects the dissipative nature of fluid membranes, as they can deform and flow locally, allowing part of the stress to dissipate without contributing to the displacement of the vesicle as a whole.

\section*{Conclusions}

Our results highlight the critical role of membrane mechanics in mediating force transduction from localized internal stresses---generated by the roto-translation of an encapsulated particle---to whole-vesicle motion. While solid-like membranes transmit particle-generated stresses efficiently to the vesicle center of mass through rigid-body rotation, fluid membranes deform and flow locally, dissipating part of the stress. This leads to stick-slip vesicle propulsion. Beyond regulating stress transmission, the fluid membrane also shapes the dynamic interplay between particle motion and vesicle deformation, which feeds back on the translation dynamics of the vesicle. The motion of the encapsulated particle is strongly influenced by the presence of boundaries outside the vesicle. Moreover, the particle is slowed down by larger separation from the membrane imposed by repulsive forces arising from membrane shape fluctuations. Simultaneously, the fluid membrane deforms in response to the stresses exerted by the encapsulated particle, redistributing tension and accommodating excess area through local curvature changes. In phase-separated membranes, line tension at the domain boundaries further influences motion by deflecting the particle from the direction imposed by the external force and gradually reorienting the domains. When the forces exerted by the particle overcome line tension, the initial Janus morphology evolves into a belt-like configuration, which acts as a guiding track to stabilize the particle trajectory along the vesicle equator. Together, these observations show that the membrane is an active mechanical component that stores, dissipates and redirects internal stresses, reshaping itself in response to persistent active forces. This dynamic coupling between particle motion and membrane remodeling may help explain how force transduction and structural organization are co-regulated in biological systems, where internal motors operate within deformable, fluid membranes.

\section*{Materials \& Methods}
\subsection*{Chemicals}
All lipids and cholesterol were purchased from Avanti Polar Lipids. Peg, PVA, hexane, chloroform, glucose, sucrose and trimethoxy(octadecyl)silane were purchased from Aldrich. BSA was purchased from Biowest. Epoxi was puchased from RS Components. Acetone, isopropanol and ethanol were purchased from VWR.
\subsection*{Microfluidic production of lipid vesicles containing ferromagnetic particles}
We use the glass microfluidic device described in \cite{arriaga_ultrathin_2014}. Briefly, two tapered capillaries with an outer diameter (OD) of 1 mm (World Precision Instruments, WPI) are positioned coaxially within a square capillary with an inner dimension (ID) of 1 mm (VitroCom, CM Scientific). The tips of the cylindrical capillaries are facing each other within the square capillary and are separated by \SI{60}{\micro\meter}. The tip diameter is \SI{60}{\micro\meter} for the injection capillary and \SI{120}{\micro\meter} for the collection capillary. The injection capillary is coated with trimethoxy(octadecyl)silane to ensure its walls are hydrophobic. Besides, inside the injection capillary we add an extra cylindrical capillary (OD = 1.00 mm, WPI), stretched with a flame until it fits inside the injection capillary. This small capillary is used to inject the inner water phase of the double emulsion which contains the solution of ferromagnetic particles. The middle oil phase is injected through the injection capillary.  The outer water phase is injected through the interstices between the injection and square capillary. The interstices of the collection and square capillary are sealed with Epoxy. As the inner water phase and middle oil phase flow through the injection capillary, they create a sequence of W/O single emulsion droplets. When these droplets reach to the tip of the injection capillary, they encounter the continuous water phase, resulting in an alternating formation of single emulsion droplets and double emulsion droplets with ultrathin middle layers. These droplets are collected via the collection capillary, which directs them to the collection medium. Upon collection, single and double emulsions are easily separated due to their density differences, with single emulsions floating and double emulsions sinking to the bottom of the collection chamber. Flow rate-controlled pumps (New Era Pump Systems, Inc.) are used to inject each phase into the device. Typical flow rates of \SI{1000}{\micro\liter}/h for the inner and middle phases, and \SI{7000}{\micro\liter}/h for the outer phase, are applied to ensure operation in the dripping regime.

\subsection*{BSA-coated substrate preparation}
Cover slides (24x50 mm, VWR) are first cleaned with soap, Milli-Q water (Millipore, Merck), ethanol, isopropanol, and acetone before coating. After cleaning, \SI{100}{\micro\liter} of a freshly prepared 10\%wt BSA solution is applied to the slide and left to sit for 15 minutes. The excess BSA is then removed by gently rinsing the slide with Milli-Q water, and any remaining water is allowed to evaporate before the slides are used.

\subsection*{Magnetic set-up and Video-Microscopy}
The magnetic set-up consists of three orthogonal pairs of Helmholtz coils aligned along the $x$, $y$, and $z$ axes coupled to an inverted bright-field and epifluorescence optical microscope (Nikon ECLIPSE Ts2R). Using a custom MATLAB code and a digital-to-analog converter DAQ (Measurement Computing USB-1208HS) we send to two of the coils a sinusoidal signal, phase-shifted by $\pi/2$. The choice of which coil pairs receive the signals depends on the desired axis of rotation for the applied magnetic field. To obtain a magnetic field strength of 10 mt we amplify the signals with 500 W amplifiers (LD Systems). An oscilloscope (Tektronix TDS 2014B) is used to monitor the voltage sent to the coils. For both bright-field and fluorescence imaging, we use a CCD camera (Grasshopper3 USB3, FLIR) with a 40X air objective (NA=0.6), Nikon S PLAN FLUOR. For fluorescent images we use a 525 nm filter. The typical frame rate for bright-field images is 10 fps, except during flicker experiments, where the frame rate is increased to 35 fps. Fluorescence images are captured at 1.53 fps. When a larger field of view is required, a 0.55X (Nikon) reduction lens is employed.

\subsection*{Particle Tracking}
Image analysis was performed using a custom MATLAB script adapted from the particle tracking algorithm originally developed by John Crocker and Eric Weeks \cite{crocker_methods_1996}. This method enables the extraction of two-dimensional trajectories by identifying the center-of-mass positions of both the particle and the vesicle. While direct measurement of the particle’s $z$-position is not possible, variations in its focus provide qualitative information about its out-of-plane motion, suggesting it follows circular paths near the vesicle equator. 

\subsection*{Flickering experiments}
To obtain the fluctuation spectrum of our lipid vesicles, we record bright-field images at the equatorial plane at a frame rate of 35 fps. We use a 40X air objective with NA=0.6. At any given time, the contour of a fluctuating vesicle can be decomposed in a series of Fourier modes, $n$, where: $r(\theta)= R_v (1 + \sum_{n=1}^\infty a_n\cos{(n\theta)}+b_n\sin{(n\theta)})$ \cite{pecreaux_refined_2004}. The fluctuation spectrum, $\langle|\xi_n|^2\rangle$, is determined by time-averaging of the quadratic fluctuation amplitudes.
\begin{equation}
\langle|\xi_n|^2\rangle=\frac{\pi\langle R_v\rangle^3}{2}[\langle|c_n|^2\rangle-\langle|c_n|\rangle^2]
\end{equation}
where $|c_n|=\sqrt{a_n^2+b_n^2}$. Following Ref. \cite{pecreaux_refined_2004}, we detect the vesicle contour, compute $a_n$, $b_n$, and $c_n$ along with their corresponding errors, and apply a correction for the finite spatial resolution of the images to account for pixelation-induced background noise. The fluctuating spectrum relates to the membrane tension, $\sigma$, and bending $\kappa$ in planar membranes through the equipartition theorem $\cite{helfrich_steric_1978}$ yielding:
\begin{equation}\label{Helfrich}
\langle|\xi_q|^2\rangle=\frac{k_BT}{(\sigma q_\perp^2 +\kappa q_\perp^4)}
\end{equation}
where $q_\perp=\sqrt{q_x^2+q_y^2}$. For vesicles only the equator is accessible and equation \ref{Helfrich} transforms to \cite{pecreaux_refined_2004}: 
\begin{equation}\label{Helfrich_ves}
\left\langle \left| \xi(q_x, y=0) \right|^2 \right\rangle = \frac{k_B T}{2 \sigma} \left[ \frac{1}{q_x} - \frac{1}{\sqrt{\frac{\sigma}{\kappa} + q_x^2}} \right]
\end{equation}
Due to the limited integration time of our camera, $\tau=30$ms equation \ref{Helfrich_ves} is no longer valid because fluctuations with a shorter life-time than $\tau$ are not correctly fitted. Therefore, for the fitting performed in Fig. \ref{flickering} we use the time corrected fluctuation spectrum proposed in \cite{pecreaux_refined_2004}:
\begin{align}\label{fit}
\left\langle \left| \xi(q_x, y = 0) \right|^2 \right\rangle & = \frac{1}{\pi} \int_{-\infty}^{\infty} \frac{k_B T}{4 \eta q_\perp} \tau_m \frac{\tau_m^2}{\tau^2}\nonumber \\
&\times\left[ \frac{\tau}{\tau_m} + \exp\left( -\frac{\tau}{\tau_m} \right) - 1 \right] \, \mathrm{d}q_y.
\end{align}
where, $\eta$ is the viscosity of the solvent and $\tau_m$ is the fluctuation life-time:
\begin{equation}
\tau_{\mathrm{m}}^{-1} = \left( \frac{1}{4 \eta q_\perp} \right) \left( \sigma q_\perp^2 + \kappa q_\perp^4 \right)
\end{equation}

\begin{acknowledgments}
The authors acknowledge financial support by MCIN/AEI/10.13039/501100011033/ for all grants listed next: LRA and JLA for PID2022-143010NB-I00 and CEX2023-001316-M, also supported by "ERDF A way of making Europe"; JLA for RYC2019-028189-I and CNS2023-145447, LRA for CNS2023-145460 also supported by "European Union NextGenerationEU/PRTR"; PM for PRE2019-091190, also supported by "ESF Investing in your future"
\end{acknowledgments}

%

\end{document}